\documentclass[aps,twocolumn,showpacs,preprintnumbers,floats]{revtex4}

\usepackage{graphicx}
\usepackage{dcolumn}
\usepackage{bm}
\usepackage{epsfig}

\begin{document}

\title{In-medium Properties of $\Theta^{+}$ as a K$\pi$N structure
 in Relativistic Mean Field Theory}
\author{
 X.\ H.\ Zhong,$^1$\footnote{E-mail: zhongxianhui@mail.nankai.edu.cn}
 G.\ X.\ Peng,$^{2,3}$\footnote{E-mail: gxpeng@ihep.ac.cn } and
 P.\ Z.\ Ning,$^{1}$\footnote {E-mail: ningpz@nankai.edu.cn}
       }
\affiliation{
 $^1$Department of Physics, Nankai University, Tianjin 300071, China \\
 $^2$Institute of High Energy Physics, Chinese Academy of Sciences,
 Beijing 100039, China\\
 $^3$China Center of Advanced Science and Technology
       (World Lab.), Beijing 100080, China           }

\begin{abstract}
The properties of nuclear matter are discussed with the
relativistic mean-field theory (RMF).
Then, we use two models in studying the in-medium properties
of $\Theta^+$: one is the point-like $\Theta^*$ in the usual RMF
and the other is a K$\pi$N structure for the pentaquark.
It is found that the in-medium properties of $\Theta^+$ are
dramatically modified by its internal structure. The effective
mass of $\Theta^+$ in medium is, at normal nuclear density,
about 1030
MeV in the point-like model, while it is
about 1120
MeV in the model of K$\pi$N pentaquark. The
nuclear potential depth of $\Theta^+$ in the K$\pi$N model is
approximately $-37.5$ MeV, much shallower than
$-90$ MeV in the usual point-like RMF model.
\end{abstract}

\maketitle

\section{Introduction}

%
The relativistic mean field theory (RMF) is one of the most
popular methods in modern nuclear physics.
%
It has been successful in describing the properties of
ordinary nuclei/nuclear matter and hyper-nuclei/nuclear matter
\cite{Rufa88prc38,Toki,Toki1,Lalazissis97prc55,Rufa90prc42,%
Mares94prc49,Schaffner96prc53,Meng,Shen,Ls,Ls1,%
Tan01CPL18,Tan04prc70,Zhong05prc71,m22,m23,m222}.
%
%
%
Appropriate effective meson-baryon interactions are essential to
the RMF calculation. To describe the nuclear matter and/or finite
nuclei, nonlinear self-interactions for $\sigma$ and $\omega$
mesons are introduced \cite{Boguta83PLB120,Bodmer91NPA526}. In
recent years, a number of effective interactions for meson-baryon
couplings, e.g., the NL-Z \cite{Rufa88prc38}, NL3
\cite{Lalazissis97prc55}, NL-SH \cite{Sharma93PLB312}, TM1, and
TM2 \cite{Sugahara94PTH92} etc., have been developed.
%

%
%
Given that RMF has been a favorite model in  describing the
properties of ordinary nuclei/nuclear matter and
hyper-nuclei/nuclear matter, we will study the in-medium
properties of $\Theta^{+}$ within the framework of the
relativistic mean field theory.

%

 The pentaquark state $\Theta^{+}$ (1540) was
first predicted by Diakonov \emph{et al.} \cite{p3}, attained much
support in the following years \cite{p4, p5, p6, p7, p8, p9, p10},
and finally listed in the review of particle physics \cite{p101}.
Presently, experimental results on $\Theta^+$ are a little subtle
(see Ref.\ \cite{n1} for a recent review), e.g., the newly
published data by the CLAS collaboration turn out to be
significantly different from the previous results \cite{n2}.
The negative results have higher statistics and are quite
convincing, but they may not completely wash away the evidence
yet, or, in other words, the pentaquark is not quite dead. Just
because of the uncertainties, it is necessary to study the
in-medium properties of $\Theta^{+}$, which is helpful to look for
signals in experiments to see whether it can exist as a bound
state in nuclei.

In fact, the study of the in-medium properties of $\Theta^{+}$ has
been a hot topic for nuclear physicists \cite{a,a1,a2,a3,a4}. Since
Miller predicted an attractive $\Theta^+$ nuclear interaction
which is strong enough to bind $\Theta^+$ in a nucleus \cite{a},
Cabrera et al. found a large attractive $\Theta^+$ potential of
$60\sim 120$ MeV in the nuclear medium.
%
%
Other investigations also show attractions, e.g., the QCD
sum-rules give an attractive $\Theta^+$ potential $40\sim 90$ MeV
\cite{a3}, and the quark mean-field model provides an attractive
potential  about $50$ MeV \cite{a4}.

In a previous paper, we have also studied the properties of
$\Theta^+$ in the nuclear medium \cite{Zhong05prc71}. However, the
internal structure of $\Theta^{+}$ was not considered there. In
fact, the internal structure is very important
\cite{m9,m10,m11,m12,m13}. It is well known that the internal
structure of $\Theta^+$ will be different if the mass of the
constituents is different. In this paper, we consider the
possibility that $\Theta^+$ is a K$\pi$N molecule state. We then
discuss a possible bound state of $\Theta^+$ in nuclei. For
comparison, we also give the results without considering the
internal structure.

The paper is organized as follows. In the subsequent section, we
present the general RMF theory for nuclear matter and for a baryon
in nuclear matter. Then the medium modifications of the pentaquark
 as a K$\pi$N structure
in medium are accordingly investigated within the framework of RMF
in Sec.\ \ref{Theta}. Finally a summary is given in Sec.\ \ref{Sum}.

\section{nuclear matter properties in RMF theory}
\label{NN}

In RMF, the effective Lagrangian density \cite{ff1,ff2,m222} can
be written as
\begin{eqnarray}
\mathcal{L}
&=&
 \sum_B
 \bigg[
  \bar{\Psi}_{B} ( i\gamma^{\mu}\partial_{\mu}
 -M_B)\Psi_B-g_{\sigma}^B\bar{\Psi }_B\sigma\Psi_B
\nonumber\\
& &
 \left.
 -g_{\omega}^B\bar{\Psi}_B\gamma^{\mu}\omega_{\mu}\Psi_B
 -g_{\rho}^{B} \bar{\psi}_{B}
  \gamma^{\mu}\rho^a_{\mu}\frac{\tau^B_a}{2}\Psi_B
 \right]
\nonumber\\
& &
 +\frac{1}{2}\partial^{\mu}\sigma\partial_{\mu}\sigma
 -\frac{1}{2}m_{\sigma}^{2}\sigma^{2}
 -\frac{1}{3}g_{2}^{2}\sigma^{3}
 -\frac{1}{4} g _{3}^{2}\sigma^{4}
\nonumber\\
& &
 -\frac{1}{4}\Omega^{\mu\nu}\Omega_{\mu\nu}
 +\frac{1}{2}m_{\omega}^{2}\omega^{\mu}\omega_{\mu}
 -\frac{1}{4}  R ^{a \mu\nu}R _{\mu\nu}^{a}
\nonumber\\
& &
 +\frac{1}{2} m_{\rho}^{2} {\rho^{a\mu}} {\rho^{a}_{\mu}}
 -\frac{1}{4} F^{\mu\nu}F_{\mu\nu}
\nonumber\\
& &
 -e\bar{\Psi }_{B}\gamma^{\mu}A^{\mu}\frac{1}{2}(1+\tau^{B}_{3})\Psi_{B}
\end{eqnarray}
with
\begin{eqnarray}
\Omega^{\mu\nu}
&=&
 \partial^{\mu}\omega^{\nu}
 -\partial^{\nu}\omega^{\mu}, \cr
 R^{a\mu\nu}
&=&
 \partial^{\mu}\rho^{a\nu}
 -\partial^{\nu}\rho^{a\mu},\cr
F^{\mu\nu}
&=&
  \partial^{\mu}A^{\nu}-\partial^{\nu}A^{\mu}.
\end{eqnarray}

The standard RMF Lagrangian involves baryons ($\Psi_B$), scalar
mesons ($\sigma$), vector mesons ($\omega_{\mu}$), vector
isovector mesons ($\rho_{\mu}$), and photons ($A_{\mu}$). The sum
on $B$ is over protons, neutrons, hyperons or exotic baryon
$\Theta^+$.

The baryon mass is $M_B$,
while the masses of $\sigma$, $\omega$, $\rho$ mesons are,
respectively, $m_{\sigma}$, $m_{\omega}$, and $m_{\rho}$.
$g_{\sigma}^{B}$, $g_{\omega}^{B}$, and $g_{\rho}^{B}$ are,
respectively, the $\sigma$-baryon, $\omega$-baryon and
$\rho$-baryon coupling constants. The Pauli matrices for baryons
are written as $\tau^B_a$ with $\tau^B_{3}$ being the third
component.

Using the mean-field approximation, i.e., replacing the meson
fields by their mean values, and neglecting the coulomb field, we
immediately have the equation of motion for baryons:
\begin{equation}
\left(
 \gamma_{\mu}k^{\mu}
 -M_B^*
 -g_{\omega}^{B}\gamma^{0}\omega_{0}
 -g_{\rho}^B\gamma^{0}\tau^{3}\rho_{03}
\right) \Psi_B
=0.
\end{equation}
where
\begin{equation} \label{mbsdef}
M^*_B
\equiv M_B+g_{\sigma}^B\sigma_0
\end{equation}
is the effective mass of baryons.
For infinite nuclear matter, the equations of motion for the
mean-field values of the scalar and vector mesons, i.e.,
$\sigma_0$ and $\omega_0$, are given by
\begin{eqnarray}
m_{\sigma}^{2}\sigma_{0}+g_{2}\sigma_{0}^2+g_{3}\sigma_{0}^3
&=&
 - g_{\sigma}^{B}\rho_{s} ,  \label{eqsig0} \\
 m_{\omega}^{2} \omega_{0} &=& g_{\omega}^{B}\rho ,
                                         \label{eqome0}
\end{eqnarray}
where $\rho_{s}$ and $\rho$ are the baryon scalar density and
vector density, respectively, which are given by
\begin{equation} \label{rhoss}
\rho_{s}\equiv \sum_{B}\langle\bar{\Psi}_{B}\Psi_{B}\rangle=
\frac{2}{(2\pi)^3}\sum_{B}\int_{0}^{k_{F}(B)}d\vec{k}
(\vec{k}^2+{M_{B}^*}^2)^{1/2},
\end{equation}
and
\begin{equation} \label{rho}
 \rho\equiv\sum_B\langle\bar{\Psi}_B\gamma^0\Psi_B\rangle
 =\sum_B\frac{{k_{\mathrm{F}}}^3(B)}{3\pi^2}.
\end{equation}

If, only one impurity baryon, e.g., $\Theta^+$, is in symmetric
infinite nuclear matter, the effect of impurity baryon on the mean
field values can be neglected \cite{f1}. Then, Eqs.\ (\ref{rhoss})
and (\ref{rho}) are simplified, giving

\begin{eqnarray}
\rho_{s} &=&\langle\bar{\Psi}_\mathrm{N}\Psi_\mathrm{N}\rangle=
 \frac{4}{(2\pi)^3}\int_0^{k_{\mathrm{F}}}\!
 \frac{M_{\mathrm{N}}^*\,\mathrm{d}\vec{k}}
      {({\vec{k}}^2+{M_{\mathrm{N}}^*}^2)^{1/2}}
\nonumber\\
&=&
 \frac{M_{\mathrm{N}}^*}{\pi^2}
 \left[
  k_{\mathrm{F}}E_{\mathrm{F}}^*
  -{M_{\mathrm{N}}^*}^2
   \ln\frac{k_{\mathrm{F}}+E_{\mathrm{F}}^*}{M_{\mathrm{N}}^*}
 \right],
\label{rhosms}
\end{eqnarray}
and
\begin{equation} \label{rho1}
 \rho =\langle
\Psi_\mathrm{N}^{\dagger}\Psi_\mathrm{N}\rangle= \frac{{2k_F}^3
}{3\pi^2},
\end{equation}
where
\begin{eqnarray}
\label{EFs}
E_{\mathrm{F}}^*
= \left({M_{\mathrm{N}}^*}^2+k_{\mathrm{F}}^2\right)^{1/2}, \ \
\label{kFN}
k_{\mathrm{F}}
=\left(\frac{3}{2}\pi^2\rho\right)^{1/3}.
\end{eqnarray}

\begin{table}[ht]
\begin{tabular}{|c|c|c|c|c|c|c|c|}\hline\hline
  & $m_{\sigma}$ & $m_{\omega}$ & $g_{\sigma }^{\mathrm{N}}$ & $g_{ \omega}^{\mathrm{N}}$
   & $g_{ \rho}^{\mathrm{N}}$ & $g_{2}$ & $g_{3}$ \\ \hline
A & 526.059 & 783.0   & 10.444 & 12.945 & 4.383 & -6.9099 & -15.8337\\
B & 508.194 & 782.501 & 10.217 & 12.868 & 4.474 & -10.434 & -28.885 \\
C & 550     & 783.0   & 9.55   & 11.67  &       &         &         \\ \hline
 \end{tabular}
\caption{Three sets of parameters.
A, B, and C are, respectively, from NL-SH, NL3, and Ref.\ \cite{b}.
The masses are given in MeV and the coupling
$g_{2}$ in fm$^{-1}$. The mass of $\rho$ mesons is $m_{\rho}=763.0$ MeV,
and the nucleon mass is $M_{\mathrm{N}}=939.0$ MeV for all the sets.}
\label{parameter}
 \end{table}

\begin{figure}[hbt]
\epsfig{file=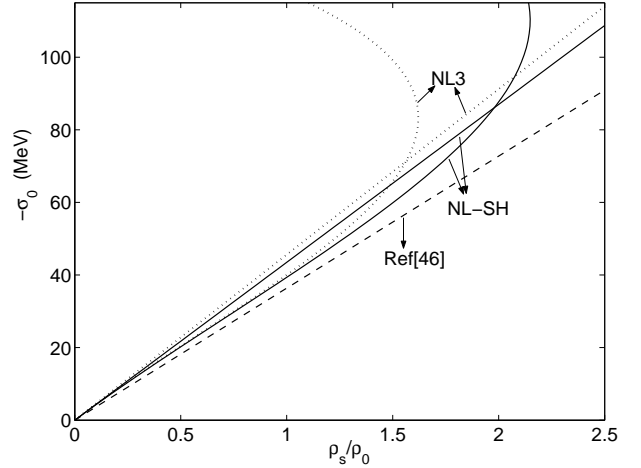,width=8.2cm}
\caption{
 The scalar meson mean-field value as a function of
 the scalar density. The straight lines are for the results
 without nonlinear $\sigma_0$ terms, and
 the curves correspond to the results with nonlinear
 $\sigma_0$ terms. The dotted line/curve is the result for
 NL3, the solid curve is for the NL-SH parameter set,
 and the dashed line is for the parameters from Ref.\ \cite{b}.
        }
 \label{sig}
\end{figure}

\subsection{The relation of $\sigma_{0}-\rho_{s}$}

Ignoring the nonlinear $\sigma_0$ terms in Eq.\ (\ref{eqsig0}),
one then has a simple linear relation between the scalar mean
field $\sigma_0$ and the scalar density $\rho_s$, i.e.,
\begin{eqnarray}
m_{\sigma}^{2} \sigma_{0}
 =g_{\sigma}^{\mathrm{N}}\rho_{s}.
\end{eqnarray}
This approximation is often used to estimate some properties of
nuclear matter. In fact, the nonlinear terms of $\sigma_{0}$ are
important to the RMF calculations, especially at dense matter.
To see clearly the effects of the nonlinear terms of $\sigma_{0}$,
we plot, in Fig.\ \ref{sig}, the mean-field value $\sigma_{0}$ as
a function of the nuclear scalar density $\rho_{s}$, with and without
nonlinear $\sigma_{0}$ terms.

In the calculations, we adopt three sets of parameters,
respectively from NL-SH, NL3, and the parameters from Ref.\
\cite{b}, which are marked with A, B, and C in Tab. I. In Fig.\
\ref{sig}, the dotted, solid, and dashed lines correspond,
respectively, to these three sets of parameters.
Obviously, the curves corresponding to the results with the
nonlinear $\sigma_{0}$ terms are different from the results
without the nonlinear terms of $\sigma_{0}$ (the straight lines).
From Fig.\ \ref{sig}, we also see the parameter dependence of the
relation $\sigma_{0}-\rho_{s}$, especially in the high-density
region.

It is interesting to note that the scalar density $ \rho_{s}$ has,
with the nonlinear $\sigma_0$ terms, an upper limit
($\rho_{s\mathrm{max}}\simeq 1.6 \rho_{0}$ for NL3 set and 2.14$
\rho_{0}$ for NL-SH set), which does not exist with the linear
relation.

\begin{figure}[ht]
\center
\epsfig{file=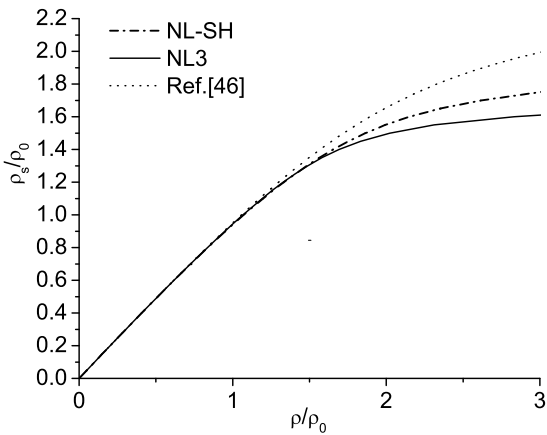,width=8.2cm}
\caption{The scalar density as a function of the nuclear density
 for three parameter sets  NL3, NL-SH, and the parameters in
 Ref.\ \cite{b}, respectively.}
 \label{rhosr}
\end{figure}

\subsection{The relation of $\rho_{s}-\rho$}

Substituting  Eq.\ (\ref{EFs}) 
into (\ref{rhosms}),
we have the following relation between the scalar density $\rho_s$,
the nucleon density $\rho$, and the $\sigma$ mean-field value $\sigma_0$:
\begin{equation} \label{rhosx}
\rho_s
= \left(
   M_{\mathrm{N}}+g_{\sigma}^{\mathrm{N}}\sigma_0
  \right)^3 f(x),
\end{equation}
where the function $f$ is defined to be
\begin{equation}
f(x)\equiv
 \left[
  x\sqrt{1+x^2}-\ln\left(1+\sqrt{1+x^2}\right)
 \right]/\pi^2
\end{equation}
with $x$ being the ratio of the nucleon's Fermi momentum to its
effective mass, i.e.,
\begin{equation} \label{xdef}
x\equiv\frac{k_{\mathrm{F}}}{M_{\mathrm{N}}^*}
=\left(\frac{3}{2}\pi^2\rho\right)^{1/3}
 \Big/\left(M_{\mathrm{N}}+g_{\sigma}^{\mathrm{N}}\sigma_0\right).
\end{equation}
The mean-field value $\sigma_0$  
is connected to the scalar density $\rho_s$ by Eq.\
(\ref{eqsig0}), i.e.,
\begin{equation} \label{rhossigma0}
\rho_s
=
-\left(
  m_{\sigma}^{2} \sigma_{0}+g_{2}\sigma_{0}^2+g_{3}\sigma_{0}^3
 \right)
/g_{\sigma}^{\mathrm{N}}.
\end{equation}
Therefore, for a given nucleon density $\rho$, we can first
solve $\sigma_0$ from
\begin{equation}
m_{\sigma}^{2} \sigma_{0}+g_{2}\sigma_{0}^2+g_{3}\sigma_{0}^3
=-g_{\sigma}^{\mathrm{N}}
 \left(
  M_{\mathrm{N}}+g_{\sigma}^{\mathrm{N}}\sigma_0
 \right)^3 f(x),
\end{equation}
and then the scalar density $\rho_s$ can be calculated
from Eq.\ (\ref{rhossigma0}) or (\ref{rhosx}).

Numerical results are shown in Fig.\ \ref{rhosr}, where the dotted
curve is the result without nonlinear $\sigma_{0}$ terms and the
parameter set is from Ref.\ \cite{b}, the solid and dash-dotted
curves are the results with the nonlinear $\sigma_{0}$ terms
corresponding, respectively, to NL3 and NL-SH. One can see, from
Fig.\ \ref{rhosr}, that the differences among the three curves are
nearly invisible in the region $\rho/\rho_{0}<1$, which indicates
that the effect from nonlinear $\sigma_{0}$ terms on the scalar
density $\rho_{s}$ is trivial, and there is little parameter
dependence for scalar density $\rho_{s}$ in the region
$\rho/\rho_{0}<1$. However, the differences become more and more
obvious with increasing densities. Thus, the nonlinear
$\sigma_{0}$ terms are more important for calculations at higher
densities, 
and in the case, the scalar density is also sensitive to the
parameter choice.

\section{ In-medium properties of $\Theta^+$}\label{Theta}

\subsection{The point-like $\Theta^+$ in RMF}

The study of the hadron properties in nuclear matter is one of the
most interesting topics in nuclear physics.
The effective mass  and nuclear potential of hadrons are two
important aspects which should be studied. In the usual RMF
framework, the effective mass of $\Theta^+$ is given by
\cite{Zhong05prc71}
\begin{eqnarray}  \label{Mstheta0}
M^{*}_{\Theta^+}
=M_{\Theta^+}+\frac{4}{3}g_{\sigma}^{\mathrm{N}}\sigma_0,
\end{eqnarray}
and the nuclear potential for $\Theta^+$ is
\begin{eqnarray} \label{Utheta0}
U_{\Theta^+} =\frac{4}{3}g_{\sigma}^{\mathrm{N}}\sigma_{0}
+\frac{4}{3}g_{\omega}^{\mathrm{N}} \omega_0.
\end{eqnarray}
According to the analysis in Sec.\ \ref{NN}, the effective
mass and nuclear potential of $\Theta^+$ as functions of $\rho$
can be obtained easily. The calculations from Eqs.\
(\ref{Mstheta0}) and (\ref{Utheta0}) will be discussed later.

\subsection{The K$\pi$N structure of $\Theta^+$ in RMF}

In the usual RMF model, the $\Theta^+$ is roughly regarded as a
point-like particle in the calculation. However,  It's very likely
that $\Theta^+$ is a bound state of quark aggregations such as
diquark-triquark $(ud\bar{s})(ud)$ \cite{m9,m10} and K$\pi$N
molecule state \cite{m11,m12,m13}. If the internal structure of
$\Theta^+$ is considered, its properties may be very different. In
the following, we will consider $\Theta^+$ as a K$\pi$N molecule
state and study its in-medium properties within the framework of
RMF.

In the picture of QMC model\cite{QMC,f1}, the mesons couple
directly with the quarks in a nucleon. 
Similarly, we can also assume that the mesons couple directly with
K, $\pi$, and N in $\Theta^+$,
given that we are considering $\Theta^+$\ as a K$\pi$N bound state.
Or, in other words, the investigation of the interaction between
$\Theta^+$\ and nucleons in medium turns into the investigation of the
interactions of KN, $\pi$N, and NN in medium. In the calculation,
we assume that the $\pi$ mesons, as Goldstone bosons, do not
change their properties in the medium \cite{E}, i.e.,
we neglect the $\pi$N interaction in medium as an approximation.

\subsubsection{KN interactions in the nuclear medium}

The NN interaction has been investigated in Sec.\ \ref{NN}. Before
investigating the in-medium properties of $\Theta^+$ with a
K$\pi$N structure, we will review the KN in-medium interaction,
which is essential to the study of $\Theta^+$ as a bound state in
the nuclear medium.

Roughly speaking, there are two popularly used methods
for the study of kaons in nuclear medium.
One is RMF theory, and the other is chiral perturbation theory
(ChPT).

\paragraph{RMF approach}

Kaons can be incorporated into the RMF model by using kaon-nucleon
interactions motivated by one meson exchange
models \cite{Schaffner96prc53,m3}.
In the meson-exchange picture, the scalar and vector interaction
between kaons and nucleons are mediated by the exchange of $\sigma$
and $\omega$ mesons. For symmetric nuclear matter, the
simplest kaon-meson interaction Lagrangian is
 \begin{eqnarray}
{\cal L}_{\mathrm{K}}
&=&
 \partial_{\mu}\bar{K}\partial^{\mu}K
 -m_{\mathrm{K}}^{2}\bar{K}K
 -g_{\sigma \mathrm{K}}m_{\mathrm{K}}\bar{K}K\sigma
\nonumber\\
& &
 -ig_{\omega\mathrm{K}}(\bar{K}\partial_{\mu}K\omega^{\mu}
 -K\partial_{\mu}\bar{K}\omega^{\mu})\nonumber\\
&&  +(g_{\omega\mathrm{K}}\omega_{\mu})^{2}\bar{K}K,
\end{eqnarray}
where $\sigma$ and $\omega^{\mu}$ are the scalar and vector fields,
respectively. $g_{\sigma\mathrm{K}}$ and $g_{\omega\mathrm{K}}$
are the coupling constants between the kaon and the scalar and vector
fields. The $\sigma$-K coupling constant is chosen from the SU(3)
relation by assuming ideal mixing \cite{Schaffner96prc53}, i.e.,
\begin{eqnarray}
2g_{\omega \mathrm{K}}=2g_{\pi\pi\rho}=6.04.
\end{eqnarray}
The K-$\omega$ coupling constant can be obtained by fitting the
experimental KN scattering length \cite{m3,m4}
\begin{eqnarray}
g_{\sigma \mathrm{K}}\approx1.9\sim2.3.
\end{eqnarray}
In the present work, we set $g_{\sigma k}=g_{\sigma}^{N}/5$, which is
in the range of $1.9\sim2.3$.

At the mean-field level, the equation of motion for kaons
is
\begin{eqnarray}
&&
\left[
 \partial_{\mu}\partial^{\mu}+m_{\mathrm{K}}^2
 +g_{\sigma\mathrm{K}}m_{\mathrm{K}}\sigma_0
\right.
\nonumber\\
&& \phantom{[}
\left.
 +2g_{\omega\mathrm{K}}\omega_{0} i\partial_0
 -(g_{\omega \mathrm{K}}\omega_{0})^2
\right] K = 0,
\end{eqnarray}
where $\sigma_{0}$ and $\omega_0$ are the mean-field value of
the scalar and  vector meson fields, respectively.
Decomposing the kaon field into plan waves, we obtain the equation
\begin{equation}
-\omega^2+\vec{k}^2+m_{\mathrm{K}}^{2}
+g_{\sigma\mathrm{K}}m_{\mathrm{K}}\sigma_0
+2g_{\omega\mathrm{K}}\omega_0 \omega
-(g_{\omega\mathrm{K}}\omega_0)^2
= 0
\end{equation}
for the kaon (anti-kaon) energy $\omega $ and the momentum $k$.
The energies of kaons and antikaons in nuclear medium are then given
by
\begin{eqnarray}
\omega_{\mathrm{K}}
=\left({m_{\mathrm{K}}^*}^2+\vec{k}^2\right)^{1/2}
 +g_{\omega\mathrm{K}}\omega_0,
\label{disper1} \\
\omega_{\bar{\mathrm{K}}}
=\left({m_{\mathrm{K}}^*}^2+\vec{k}^2\right)^{1/2}
 -g_{\omega\mathrm{K}}\omega_0,
\label{disper2}
\end{eqnarray}
where the effective  mass of kaons is
\begin{eqnarray} \label{mksdef}
m_{\mathrm{K}}^*
=\sqrt{m_{\mathrm{K}}^2
 +g_{\sigma\mathrm{K}}m_{\mathrm{K}}\sigma_0}.
\end{eqnarray}
From the in-medium dispersion relations (\ref{disper1})
and (\ref{disper2}), the kaon/antikaon potential can be
defined as \cite{m5,m6}
\begin{eqnarray}
U_{\mathrm{K}/\bar{\mathrm{K}}}
=\omega_{\mathrm{K}/\bar{\mathrm{K}}}
 -\sqrt{m_{\mathrm{K}}^2+\vec{k}^2}.
\end{eqnarray}

\paragraph{Chiral approach}

For comparison, we also introduce another method, the chiral
perturbation approach in our calculations. For symmetric
nuclear matter, the effect of isospin is neglected. Following
Refs.\ \cite{m3,m7}, the kaon-nucleon chiral Lagrangian
is written as
\begin{eqnarray}
{\cal L}_{\mathrm{KN}}^{\mathrm{chiral}}
&=&
 -\frac{3i}{8f_{\mathrm{K}}^2}\bar{\Psi}_\mathrm{N}\gamma^{\mu}\Psi_\mathrm{N}
 \left[
  \bar{K}\partial^{\mu}K-(\partial^{\mu}\bar{K})K
 \right]
\nonumber\\
& &
  +\frac{\Sigma_{\mathrm{KN}}}{f_{\mathrm{K}}^2}\bar{\Psi}_\mathrm{N}\Psi_\mathrm{N}\bar{K}K\nonumber\\
& &
  +\frac{\tilde{D}}{f_{\mathrm{K}}^2}\bar{\Psi}_\mathrm{N}\Psi_\mathrm{N}(\partial_{\mu}\bar{K}\partial^{\mu}K),
\end{eqnarray}
where $f_{\mathrm{K}} \approx 93$ MeV is the kaon decay constant
and $\Sigma_{\mathrm{KN}}$ is the KN sigma term. The first term
corresponds to the Tomozawa-Weinberg vector interaction. The
second term is the scalar interaction which will shift the effective mass
of the kaon and the antikaon. The last term, which is sometimes
called the off-shell term, modifies the scalar interaction.
$\Sigma_{\mathrm{KN}}$ is not known very well, in the original
work, it was chose to be $\Sigma_{\mathrm{KN}}\approx2m_{\pi}$ in
accordance with the Born model \cite{m7}. More recently, the value
$\Sigma_{\mathrm{KN}}\approx 450\pm 30$ MeV is favored according to
lattice gauge calculations \cite{m8}. Thus, it may vary in the
region from 270 MeV to 480 MeV. By fitting the KN scattering
lengths, one can determine the constant \cite{m3}
\begin{eqnarray}
\tilde{D}=0.33/m_{\mathrm{K}}-\Sigma_{\mathrm{KN}}/m_{\mathrm{K}}^2.
\end{eqnarray}
The equation of motion for kaon field in the mean-field
approximation and in uniform matter reads
\begin{eqnarray}
&& \left[\partial_{\mu} \partial^{\mu}+m_{\mathrm{K}}^{2}
-\frac{\Sigma_{\mathrm{KN}}}{f_{\mathrm{K}}^2}\rho_{s} \right.
\nonumber\\
&& \phantom{[}
 \left.
  +\frac{\tilde{D}}{f_{\mathrm{K}}^2}\rho_{s}\partial_{\mu} \partial^{\mu}
  +\frac{3i}{4f_{\mathrm{K}}^2}\rho_{N} \partial_{0}
\right] K=0,
\end{eqnarray}
where $\rho_s=\langle\bar{\Psi}_\mathrm{N}\Psi_\mathrm{N}\rangle$
is the scalar density and $\rho=\langle
\Psi_\mathrm{N}^{\dagger}\Psi_\mathrm{N}\rangle$ is the vector
density for nucleons. Plane wave decomposition of the equation of
motion yields
\begin{eqnarray}
& &
-\omega^2+\vec{k}^2+m_{\mathrm{K}}^2
-\frac{\Sigma_{\mathrm{KN}}}{f_{\mathrm{K}}^2}\rho_s
\nonumber\\
& & \phantom{-}
 -\frac{\tilde{D}}{f_{\mathrm{K}}^2}\rho_{s}
  \left(\omega^2-\vec{k}^2\right)
 -\frac{3}{4f_{\mathrm{K}}^2}\omega\rho  =0.
\end{eqnarray}
The kaon and antikaon energies in the nuclear medium are
\begin{eqnarray}
\omega_{\mathrm{K}}
&=&
 \left\{
  \left[
   \left({m_{\mathrm{K}}^*}^2+\vec{k}^2\right)
   \left(1+\frac{\tilde{D}}{f_{\mathrm{K}}^2}\rho_{s}\right)^2
   +\left(\frac{3}{8f_{\mathrm{K}}^2} \rho \right)^2
  \right]^{1/2}
 \right.
\nonumber\\
& & \phantom{\Bigg[}
  +\frac{3}{8f_{\mathrm{K}}^2}\rho
 \Bigg\}
 \left(
  1+\frac{\tilde{D}}{f_{\mathrm{K}}^2}\rho_s
 \right)^{-1}, \\
\omega_{\bar{K}}
&=&
 \left\{
  \left[
   \left({m_{\mathrm{K}}^*}^2+\vec{k}^2\right)
   \left(1+\frac{\tilde{D}}{f_{\mathrm{K}}^2}\rho_s\right)^2
   +(\frac{3}{8f_{\mathrm{K}}^2} \rho )^2
  \right]^{1/2}
 \right.
\nonumber\\
& & \phantom{[}
 \left.
  -\frac{3}{8f_{\mathrm{K}}^2}
  \rho
 \right\}
 \left(
  1+\frac{\tilde{D}}{f_{\mathrm{K}}^2}\rho_s
 \right)^{-1},
\end{eqnarray}
where $m_{\mathrm{K}}^*$ is the kaon effective mass
\begin{eqnarray}
m_{\mathrm{K}}^*
=\sqrt{
  \left(
   {m_{\mathrm{K}}^*}^2
   -\frac{\Sigma_{\mathrm{KN}}}{f_k^2}\rho_s
  \right)
   \bigg/
  \left(
   1+\frac{\tilde{D}}{f_{\mathrm{K}}^2}\rho_s
  \right)
      }.
\end{eqnarray}

\paragraph{Results}

\begin{figure}[ht]
\center\epsfxsize=8.2 cm \epsfbox{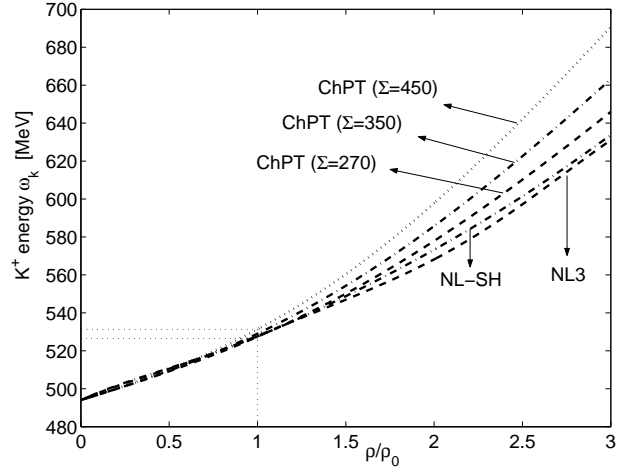} \caption{ K$^+$
energy as a function of the baryon density for RMF model and ChPT
model. For ChPT model, the results with $\Sigma_{\mathrm{KN}}=270, 350, 450$ MeV
are plotted, and for RMF model, the results with NL3 and Nl-SH
plotted.} \label{kmass}
\end{figure}

The kaon energy $\omega_{\mathrm{K}}$ at zero momentum ($k=0$) as
a function of the nuclear density $\rho$ is plotted in Fig.\
\ref{kmass}. For RMF approach, we adopt two sets of parameters:
NL-SH and NL3. For ChPT method, we consider also three cases:
$\Sigma_{\mathrm{KN}}=270, 350, 450$ MeV, and in the calculation
we have used the relation $\rho_s-\rho$ in RMF model with the
NL-SH parameter set.

From Fig.\ \ref{kmass}, we hardly see  any difference between all
the curves in the low density region $\rho<\rho_0$. And in the
region of $\rho>\rho_0$, the ChPT approach always gives larger
kaon energy than RMF model. At $\rho=\rho_{0}$, the kaon energy is
about $\omega_{\mathrm{K}}=527$ MeV, which is nearly independent
of the models and parameter sets. The kaon potential
$U_{\mathrm{K}}=33$ MeV for RMF model and $U_{\mathrm{K}}=28\sim
33$ MeV for ChPT model at normal nuclear density, compatible with
the prediction $U_{\mathrm{opt}}\approx 29$ MeV in Ref.\
\cite{m3}. At high densities, the uncertainty of the results from
different parameter sets within RMF model is $\sim 10$ MeV. At
$\rho=3\rho_0$, the kaon energy $\omega_{\mathrm{K}}=630-640$ MeV
($\omega_{\mathrm{K}}=630$ MeV for NL-Z in Ref.\ \cite{m3}) for
RMF model, and $\omega_{\mathrm{K}}=645\sim690$ MeV
($\omega_{\mathrm{K}}=630\sim 670$ MeV in Ref.\ \cite{m3}) for
ChPT model. All the results are compatible with the calculation in
Ref.\ \cite{m3} as a whole. Thus, they give us reliable bases for
the following calculations and discussions.

\subsubsection{Formulas for $\Theta^+$ with a K$\pi$N Structure in RMF}

Now that the NN and KN interactions in nuclear matter have been
known, it is convenient to investigate the K$\pi$N structure
of $\Theta^+$ in the nuclear medium. The effective
Lagrangian for $\Theta^+$ as a K$\pi$N bound state can be
written as
\begin{eqnarray}
{\cal L}_{\Theta^+}
 \simeq {\cal L}_{\mathrm{K}}
 +{\cal L}_{\mathrm{N}}
 +{\cal L}_{\pi}
 +{\cal L}_0,
\end{eqnarray}
where
${\cal L}_{\mathrm{K}}$,
${\cal L}_{\mathrm{N}}$
and ${\cal L}_{\pi}$
are the effective Lagrangian densities, respectively,
for nucleons, kaons, and pions,
and  ${\cal L}_{0}$ is for the internal interaction
of K$\pi$N system.
Usually, the contributions from ${\cal L}_{\pi}$ and ${\cal L}_0$
are small. Therefor, we do not explicitly consider them in the
following.

In the RMF framework, the interactions between hadrons are
mediated by the exchange of the scalar meson $\sigma$ and vector
meson $\omega$. According to the above discussions,
${\cal L}_{\mathrm{N}}$ and ${\cal L}_{\mathrm{K}}$
for symmetric nuclear matter can be written as
\begin{eqnarray}
{\cal L}_{\mathrm{N}}
&=&
 \bar{\Psi}_{\mathrm{N}}
  \left(i\gamma^{\mu}\partial_{\mu}-M_{\mathrm{N}}\right)\Psi_{\mathrm{N}}
 -g_{\sigma}^{\mathrm{N}}\bar{\Psi}_{\mathrm{N}}\sigma\Psi_{\mathrm{N}}
\nonumber\\
& &
 -g_{\omega}^{\mathrm{N}}\bar{\Psi}_{\mathrm{N}}\gamma^{\mu}\omega_{\mu}\Psi_{N},\\
{\cal L}_{\mathrm{K}}
&=&
 \partial_{\mu}\bar{K}\partial^{\mu}K-m_{\mathrm{K}}^2\bar{K}K
 -g_{\sigma\mathrm{K}}m_{\mathrm{K}}\bar{K}K\sigma
\nonumber\\
& &
 -ig_{\omega\mathrm{K}}
  \left(
   \bar{K}\partial_{\mu}K\omega^{\mu}
   -K\partial_{\mu}\bar{K}\omega^{\mu}
  \right)
\nonumber\\
& &
 +(g_{\omega\mathrm{K}}\omega_{\mu})^2\bar{K}K.
\end{eqnarray}
Then, at the mean-field level, the in-medium energy of nucleons is
given by
\begin{eqnarray}
E_{\mathrm{N}}(\vec{p}) =\sqrt{
  \left(
   M_{\mathrm{N}}+g_{\sigma}^{\mathrm{N}}\sigma_0
  \right)^2 +\vec{p}^2
      }
 +g_{\omega}^{\mathrm{N}}\omega_0,
\end{eqnarray}
and the in-medium energy of $\Theta^+$ is
\begin{eqnarray}
E_{\Theta^+} =E_{\mathrm{N}}(\vec{p})+
\omega_{\mathrm{K}}+E_{\pi}+E_{\mathrm{b}},
\end{eqnarray}
where $E_{\pi}=\sqrt{m_{\pi}^2+\vec{p}^2}$ is the $\pi$\ energy
and $E_{\mathrm{b}}$ is the bound energy of K$\pi$N system,
neither changes in nuclear medium. At zero momentum, the in-medium
$\Theta^+$ energy is given by
\begin{eqnarray}
E_{\Theta^+}
=M_{\Theta^+}^*
 +g_{\omega}^{\mathrm{N}}\omega_0
 +g_{\omega\mathrm{K}}\omega_0,
\end{eqnarray}
where the effective mass of $\Theta^+$\ in nuclear matter is
\begin{eqnarray}
M_{\Theta^+}^*=M_{\mathrm{N}}^*+m_{\mathrm{K}}^*+m_{\pi}+E_{\mathrm{b}}.
\end{eqnarray}
In free space, the $\Theta^+$ mass is
$M_{\Theta^+}=M_{\mathrm{N}}+m_{\mathrm{K}}+m_{\pi}+E_{\mathrm{b}}$.
Combining Eqs.\ (\ref{mbsdef}) and (\ref{mksdef}), the effective
mass of $\Theta^+$ can be expressed as
\begin{eqnarray} \label{Mthetakpin}
M_{\Theta^+}^*
=M_{\Theta^+}
 +g_{\sigma}^{\mathrm{N}}\sigma_0
 +\sqrt{m_{\mathrm{K}}^2
 +g_{\sigma\mathrm{K}}m_{\mathrm{K}}\sigma_0}
 -m_{\mathrm{K}}.
\end{eqnarray}

The $\Theta^+$ potential is defined to be the energy difference
between $\Theta^+$ in the medium and $\Theta^+$ in free space,
i.e.,
\begin{equation}
U_{\Theta}^{\mathrm{K}\pi\mathrm{N}}(\vec{p},\vec{k})
=E_{\mathrm{N}}(\vec{p})+\omega_{\mathrm{K}}
 -\sqrt{M_{\mathrm{N}}^2+\vec{p}^2}
 -\sqrt{m_{\mathrm{K}}^2+\vec{k}^2}.
\end{equation}

At zero momentum ($p=k=0$), one then has
\begin{eqnarray}
U_{\Theta}^{\mathrm{K}\pi\mathrm{N}}
&=&
 g_{\sigma}^{\mathrm{N}}\sigma_0
 +g_{\omega}^{\mathrm{N}} \omega_0
 +\left[
   m_{\mathrm{K}}^2
   +g_{\sigma\mathrm{K}}m_{\mathrm{K}}\sigma_0
  \right]^{1/2}
\nonumber\\
& &
 +g_{\omega\mathrm{K}}\omega_{0}-m_{\mathrm{K}}.
\end{eqnarray}

\begin{figure}[ht]
\center\epsfxsize=8.2 cm \epsfbox{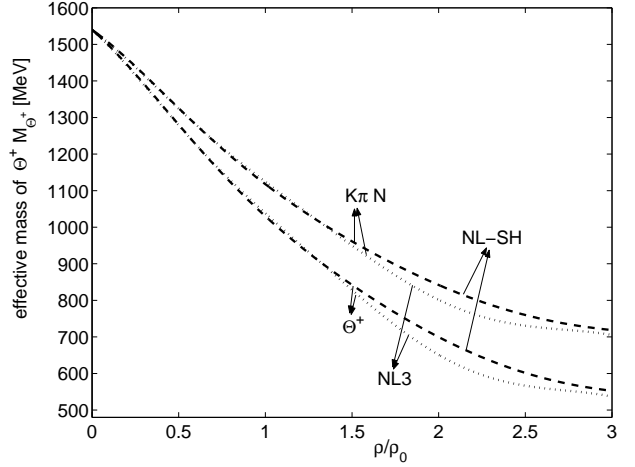} \caption{ The
effective mass of $\Theta^{+}$ with two models in nuclear medium.
One is the $\Theta^{+}$ as a point-like particle (denoted with
$\Theta^{+}$), the other is $\Theta^{+}$ as a K$\pi$N structure
(denoted with K$\pi$N). The results with parameter sets NL-SH and
NL3 are plotted. } \label{emass}
\end{figure}

\subsection{Effective mass of $\Theta^{+}$ in RMF}

Now we can easily obtain the effective masses, respectively, for
the point-like $\Theta^+$ and for the $\Theta^+$ with a K$\pi$N
structure, according to the Eqs.\ (\ref{Mstheta0}) and
(\ref{Mthetakpin}). The results are given in Fig.\ \ref{emass}.
Two parameter sets, i.e., the NL-SH and NL3 are adopted in the
calculations. It is found that, when $\Theta^+$ is regarded as a
K$\pi$N bound state, its effective mass is dramatically enhanced.
At normal nuclear density, the effective mass of K$\pi$N system is
$M_{\Theta^{+}}^*\simeq 0.73 M_{\Theta^+}=1120$ MeV, much larger
than that of the point-like $\Theta^+$ ($M^*_{\Theta^+}\simeq 0.67
M_{\Theta^{+}}=1030$ MeV). The effective mass of $\Theta^+$ is
enhanced by about 90 MeV (6\% of $M_{\Theta^+}$) due to its
internal structure. At higher densities, say $\rho=3\rho_0$, the
effective mass of $\Theta^+$ with K$\pi$N structure is enhanced
about 160 MeV compared with that of point-like $\Theta^+$.

The effective mass depends obviously on the parameter sets in the
region of $\rho>1.5\rho_0$, the mass difference is about $10\sim
30$ MeV between the NL-SH and NL3 parameter sets. However, at lower densities
$\rho<1.5\rho_{0}$, the difference is almost invisible between the
two curves (NL-SH and NL3).

As a whole, the effective mass of $\Theta^+$ depends strongly on
its internal structure. The difference is up to 90 MeV between
K$\pi$N structure and point-like $\Theta^+$ at $\rho=\rho_0$.
The effects on the effective mass for different parameter sets
of the point-like $\Theta^+$ are much smaller than those of
the K$\pi$N system.

\subsection{Nuclear potential depth of $\Theta^{+}$ in RMF}

\begin{figure}[ht]
\center\epsfxsize=8.2 cm \epsfbox{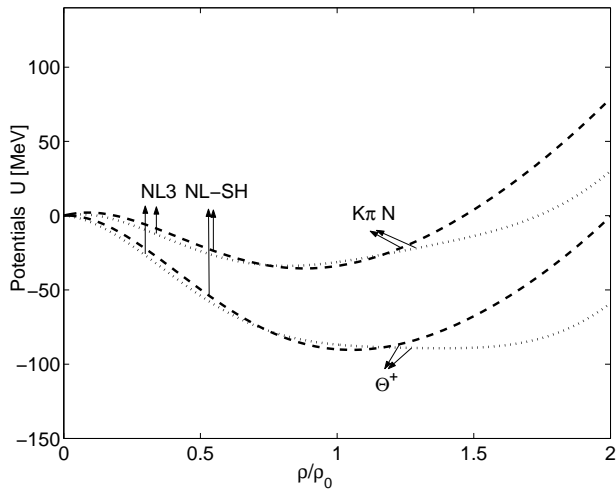} \caption{ The
potentials of $\Theta^{+}$ with two models in nuclear medium. One
is the $\Theta^{+}$ as a point-like particle (denoted with
$\Theta^{+}$), the other is $\Theta^{+}$ as a K$\pi$N structure
(denoted with K$\pi$N). The results with parameter sets NL-SH and
NL3 are plotted.} \label{pp}
\end{figure}

The potential depth is another important aspect for understanding
the interactions between $\Theta^{+}$ and nucleons. To see the
$\Theta^+$ potential in nuclear matter clearly, the $\Theta^+$
potential of the K$\pi$N structure is plotted in Fig.\ \ref{pp}.
For comparison, the potential of the point-like $\Theta^{+}$ as a
function of the nuclear density $\rho$ is also plotted in the same
figure.

From the figure, we can see that when $\Theta^{+}$ is considered
as a K$\pi$N bound state, the potential depth becomes much
shallower than that of point-like $\Theta^+$. At normal nuclear
density, the potential depth of $\Theta^+$ with K$\pi$N structure
is about $-37.5$ MeV, which is about 52 MeV shallower than that of
the point-like $\Theta^{+}$. The potential depth of $\Theta^{+}$
with K$\pi$N structure at normal nuclear density is also shallower
than the previous predictions in other models \cite{a2,a3,a4}.
Therefore, if $\Theta^+$ is indeed a K$\pi$N bound state, it
should have a shallower in-medium potential in RMF. This is a
quite different observation from the previous result of strong
binding (several hundreds of MeV) in nuclei \cite{a,a2}.

On the other hand, we see, from the figure, that the results
for NL3 and NL-SH are very similar in the region $\rho<1.2
\rho_{0}$. However, there are strong parameters dependence in the higher
density region $\rho>1.2\rho_{0}$.

\section{Summary}\label{Sum}

Considering the $\Theta^+$ as a K$\pi$N structure,
we have calculated the effective mass and optical potential
of $\Theta^+$\ in the nuclear medium with both
the RMF approach and the ChPT theory. We find that
the potential depth is only $U_{\Theta^+} \simeq-37.5$ MeV
at normal nuclear density, much
shallower than $U_{\Theta^+} \simeq-90$ MeV for point-like
$\Theta^+$. The effective mass of $\Theta^+$ with a K$\pi$N
structure is also dramatically enhanced, which is
$M^*_{\Theta^+}\simeq 0.73 M_{\Theta^+}=1120$ MeV at normal
density, 90 MeV larger than $M^*_{\Theta^+}\simeq 0.67
M_{\Theta^+}=1030$ MeV for point-like $\Theta^+$.

 Of course, our results depend on parameters.
Also, $\Theta^+$ may have other internal structures.
Because different internal structure have different
in-medium properties, further studies on the $\Theta^+$\
in nuclear medium are needed to clarify uncertainties.

\section*{Acknowledgements }

The authors thank support from
the Natural Science Foundation of China
(10275037, 10375074, 90203004)
and the Doctoral Programme Foundation of the China
Institution of Higher Education (20010055012).

\end{document}